\documentclass[letterpaper]{JHEP3}

\usepackage{amsmath}
\usepackage{amsthm}

\newcommand{\be}{\begin{equation}}
\newcommand{\ee}{\end{equation}}
\newcommand{\bea}{\begin{eqnarray}}
\newcommand{\eea}{\end{eqnarray}}

\newcommand{\eq}[1]{(\ref{eq:#1})}
\newcommand{\sect}[1]{Sec.~\ref{sec:#1}}
\newcommand{\append}[1]{App.~\ref{sec:#1}}

\newcommand{\tabl}[1]{Table~\ref{table:#1}}
\newcommand{\thm}[1]{Theorem~\ref{thm:#1}}

\newcommand{\ket}[1]{\mbox{$| #1 \rangle$}}
\newcommand{\norm}[2]{\mbox{$\langle #1 | #2 \rangle$}}
\newcommand{\normb}[1]{\mbox{$\parallel #1 \parallel^2$}}
\newcommand{\nor}{\stackrel{\scriptscriptstyle\circ}{\scriptscriptstyle\circ}}

\newcommand{\sig}[1]{\mbox{sign}(#1)}
\newcommand{\dims}[1]{\mbox{dim}(#1)}
\newcommand{\tr}{\mbox{tr}\,}

\newcommand{\cond}[1]{\quad\mbox{#1}}

\newcommand{\Htotal}{{\cal H}_{\rm total}}
\newcommand{\Hc}{\hat{\cal H}_{\rm c}}
\newcommand{\He}{\hat{\cal H}_{\rm e}}
\newcommand{\obs}{\rm obs}

\newcommand{\hN}{\hat{N}^g}
\newcommand{\hQ}{\hat{Q}}

\newcommand{\hX}{h^{\rm 0}}
\newcommand{\hK}{h^{K}}
\newcommand{\hKp}{h}

\newcommand{\cX}{c^0}
\newcommand{\cK}{c^{K}}
\newcommand{\Lint}{L_{0}^{\rm osc}}
\newcommand{\Lm}[1]{L^{\rm m}_{#1}}
\newcommand{\Lt}[1]{L^{0}_{#1}}
\newcommand{\tLm}[1]{\tilde{L}^{\rm m}_{#1}}
\newcommand{\tLt}[1]{\tilde{L}^{0}_{#1}}
\newcommand{\Lg}[1]{L^{g}_{#1}}
\newcommand{\LK}[1]{L^{K}_{#1}}

\newcommand{\km}[2]{J^{#1}_{#2}}
\newcommand{\tkm}[2]{{\tilde J}^{#1}_{#2}}
\newcommand{\tj}{\tilde{j}}

\newcommand{\alp}{\alpha'}

\newtheorem{theorem}{Theorem}[section]

\def\IR{\relax{\rm I\kern-.18em R}}

\title{The BRST quantization and the no-ghost theorem for $AdS_3$}

\author{Masako Asano\\
Theory Division\\
Institute of Particle and Nuclear Studies\\
KEK, High Energy Accelerator Research Organization\\
Tsukuba, Ibaraki, 305-0801 Japan\\
\email{asano@post.kek.jp}
}

\author{Makoto Natsuume
\thanks{On leave of absence from KEK.}\\
Department of Physics and Astronomy\\
University of Pennsylvania\\
Philadelphia, PA 19104-6396, USA\\
\email{natsuume@physics.upenn.edu}
}

\abstract{In our previous papers, we prove the no-ghost theorem without light-cone directions (hep-th/0005002, hep-th/0303051). We point out that our results are valid for more general backgrounds. In particular, we prove the no-ghost theorem for $AdS_3$ in the context of the BRST quantization (with the standard restriction on the spin). We compare our BRST proof with the OCQ proof and establish the BRST-OCQ equivalence for $AdS_3$. The key in both approaches lies in the certain structure of the matter Hilbert space as a product of two Verma modules. We also present the no-ghost theorem in the most general form.}

\preprint{
KEK-TH-883, UPR-1041-T \\ 
hep-th/0304254
}

\begin{document}


\section{Introduction}

Even though strings on curved backgrounds have been widely discussed in the last decade or so, rigorous discussion within string theory is very difficult. This is due to the lack of the string theory on general backgrounds, especially the no-ghost theorem. As is well-known, string theory generally contains negative norm states (ghosts) from timelike oscillators. However, they do not appear as physical states. This is well-established for string theory in flat spacetime. When the background spacetime is curved, things are not clear though. 
Standard proofs of the no-ghost theorem requires light-cone directions; if the background is written as $ \IR^{1,d-1} \times K $ with a unitary CFT $K$, $d \geq 2$. This is true both in the old covariant quantization (OCQ) and in the BRST quantization.

In our previous papers, we show the no-ghost theorem for $d \geq 1$ using the BRST quantization
\cite{Asano:2000fp}. (See Ref.~\cite{Asano:2003jn} for the NSR string). Here, we extend our results to more general backgrounds. We point out that 
\begin{itemize}
\item The vanishing theorem is valid 
if the matter Hilbert space is written as a direct product of two Verma modules (as in \thm{vanishing}), one for a nonunitary $c=1$ CFT and another for the rest, which is assumed to be a unitary CFT.
\item The no-ghost theorem is also valid for the above Hilbert space (under a certain condition. See~\thm{no-ghost}.)
\end{itemize}
One particular example is $AdS_3$, so we establish the no-ghost theorem in the BRST quantization of $AdS_3$. We heavily use the previous results on the old covariant quantization of $AdS_3$ \cite{Dixon:1989cg}-\cite{Pakman:2003cu}. In particular, we impose the restriction on the $SL(2,\IR)$ spin $j$ for the discrete series representations first proposed in Refs.~\cite{Mohammedi:1989dp,Petropoulos:1989fc,Hwang:1990aq}, and we include the spectral flowed sectors first proposed in Ref.~\cite{Henningson:1991jc}.

In the context of the OCQ, the no-ghost theorem for $AdS_3$ has been known \cite{Hwang:1990aq,Hwang:1991an,Evans:1998qu,Maldacena:2000hw}. However, the foundation of the perturbative string theory lies in the BRST quantization, and the OCQ often needs to be justified from the underlying BRST quantization. We compare our proof with the proof in the OCQ and also establish the BRST-OCQ equivalence for $AdS_3$. It turns out that the key in both approaches is the certain structure of the Hilbert space mentioned above.[Eq.~\eq{hilbert} for the BRST quantization and Eq.~\eq{OCQ_basis2} for the OCQ]

The organization of the present paper is as follows. First, in the next section, we briefly review string theory on $AdS_3$. The full proof of the no-ghost theorem is rather involved, so we give the outline of the proof in \sect{outline}. Section~\ref{sec:vanishing} and \ref{sec:no-ghost:ads3} are the discussion of the vanishing theorem and the no-ghost theorem for $AdS_3$, respectively. In \sect{no-ghost}, we present the no-ghost theorem in the most general form. We also discuss a time-dependent background of Ref.~\cite{Balasubramanian:2002ry} where the background has ghosts in certain cases and explain how the background violates the assumptions of our theorem in those cases. In \sect{equiv}, we establish the BRST-OCQ equivalence for the backgrounds our no-ghost theorem applies. We summarize our notations and conventions of the BRST quantization in \append{appA}. The presentation of the proof in our second paper \cite{Asano:2003jn} is slightly different from our first paper \cite{Asano:2000fp} although the proof itself is very similar; we will follow the style of our second paper.

\section{Strings on $AdS_3$}\label{sec:ads3}

$AdS_3$ with NS-NS flux is described by the $SL(2,\IR)$ WZW model. In most applications, one tensors an internal CFT $K'$ and considers string theory on $SL(2,\IR) \times K'$. We follow the notations and the conventions of Ref.~\cite{Maldacena:2000hw} with some minor changes. 

The Kac-Moody algebra for $sl(2,\IR)$ is
\be
[\km{a}{m}, \km{b}{n}] = i \epsilon^{ab}_{~~c} \km{c}{m+n} + \frac{k}{2}m \eta^{ab} \delta_{m+n},
\ee
where $k$ is the level, $\epsilon^{123}=+1$ and $\eta^{ab}=\mbox{diag}(+1,+1,-1)$. In terms of the modes $\km{\pm}{n}=\km{1}{n} \pm i\km{2}{n}$, the algebra is
\bea
&&[\km{+}{m}, \km{-}{n}] = -2 \km{3}{m+n} + km \delta_{m+n},\\
&&[\km{\pm}{m}, \km{\pm}{n}] = 0,\\
&&[\km{3}{m}, \km{\pm}{n}] = \pm \km{\pm}{m+n},\\
&&[\km{3}{m}, \km{3}{n}] = -\frac{k}{2} m \delta_{m+n}.
\eea
The matter Virasoro generator 
\bea
\Lm{m} &=& \frac{1}{k-2} \sum_n \eta_{ab} \nor \km{a}{m-n} \km{b}{n} \nor 
\nonumber \\
&=& \frac{1}{k-2} \sum_n \nor \left\{
\frac{1}{2} (\km{+}{m-n} \km{-}{n} + \km{-}{m-n} \km{+}{n}) - \km{3}{m-n} \km{3}{n} \right\} \nor 
\eea
satisfies the Virasoro algebra with central charge
\be
c_{SL(2,\IR)} = \frac{3k}{k-2}
\ee
and
\bea
&& [\km{\pm}{m}, \Lm{n}]=m \km{\pm}{m+n}, \\
&& [\km{3}{m}, \Lm{n}]=m \km{3}{m+n}.
\eea
We always assume $k>2$ to ensure a positive central charge as well as a single timelike direction. When there is no internal CFT $K'$, $c_{SL(2,\IR)}=26$ so that $k=52/23$.

\subsection{Unflowed representations}\label{sec:unflowed}

If the spectrum is bounded below, acting repeatedly with $\km{a}{n} \; (n>0)$ always produces a Kac-Moody primary which is annihilated by $\km{a}{n} \; (n>0)$. For the moment, let us suppose that this is the case and call these representations as ``unflowed representations" (for the reason which will soon become clear.) For the WZW model, a Kac-Moody primary is also a Virasoro primary. A Kac-Moody primary forms a representation of the global $sl(2,\IR)$ generated by the zero modes $\km{a}{0}$. Then, the representations of the $sl(2,\IR)$ current algebra are built over Kac-Moody primaries by applying $\km{a}{-n} \; (n>0)$ in all possible ways. There are five classes of the unitary representations of the global $sl(2,\IR)$. They are characterized by the second Casimir $c_2 = \eta_{ab} \km{a}{0} \km{b}{0} = -j(j-1)$ and $\km{3}{0}=m$:

\begin{enumerate}

\item Lowest weight discrete series:
\be
{\cal D}^+_j = \{ \ket{j,m}: m=j, j+1, j+2, \cdots \},
\ee
where $j>0$ such that $\km{-}{0} \ket{j,j} =0$.

\item Highest weight discrete series:
\be
{\cal D}^-_j = \{ \ket{j,m}: m=-j, -j-1, -j-2, \cdots \},
\ee
where $j>0$ such that $\km{+}{0} \ket{j,-j} =0$.

\item Principal continuous series:
\be
{\cal C}^\alpha_j = \{ \ket{j,\alpha,m}: m=\alpha, \alpha \pm 1, \alpha \pm 2, \cdots \},
\ee
where $0 \leq \alpha <1$ and $j=1/2+is$, $s>0$.

\item Complementary (Supplementary) series:
\be
{\cal E}^\alpha_j = \{ \ket{j,\alpha,m}: m=\alpha, \alpha \pm 1, \alpha \pm 2, \cdots \},
\ee
where $0 \leq \alpha <1$, $1/2<j<1$ and $j-1/2<|\alpha-1/2|$.

\item Identity representation: trivial representation with $j=0$.

\end{enumerate}

\noindent
We always consider the universal cover of $SL(2,\IR)$;
so, the spin $j$ is not restricted to be a half-integer or an integer. A complete basis for the square integrable functions on $SL(2,\IR)$ is known from the harmonic analysis; they are given by the matrix elements of the first three representations ${\cal D}^{\pm}_j$ (with $j>1/2$) and ${\cal C}^\alpha_j$. Thus, we consider only those representations. We denote the representations of the full current algebra built over those zero mode representations by ${\hat{\cal D}}^\pm_j$ and ${\hat{\cal C}}^\alpha_j$.

Now, the on-shell condition at grade $N$ reduces to 
\be
\frac{-j(j-1)}{k-2}+N+\hKp=1.
\label{eq:on-shell}
\ee
Here, $\hKp \geq 0$ is a conformal weight from the internal CFT $K'$. For the continuous series, this condition is satisfied only for $N=0$. By construction, they satisfy the other physical state conditions and they are unitary. Thus, only the discrete series are usually considered for the no-ghost theorem (for the unflowed representations.)

The no-ghost theorem for the background has been widely discussed using the OCQ \cite{Balog:1988jb}-\cite{Pakman:2003cu}. 
In particular, Refs.~\cite{Hwang:1990aq,Hwang:1991an,Evans:1998qu} proved the theorem for ${\hat{\cal D}}^\pm_j$ with the additional restriction on the spin $j$ for ${\cal D}^{\pm}_j$:
\be
0<j<\frac{k}{2}.
\label{eq:bound1}
\ee
We sketch the proof in \append{appC}. 
The bound in turn implies that the grade is bounded above from the on-shell condition \eq{on-shell}:
\be
N+\hKp<1+\frac{k}{4}
\label{eq:bound_for_grade}
\ee

\subsection{Flowed representations}

It is proposed in Ref.~\cite{Henningson:1991jc} that these representations appeared in the last subsection are not the only representations appear in the WZW model. This idea was then revived in Ref.~\cite{Maldacena:2000hw} and obtained physical significance. To see this, note that the current algebra is invariant under the following transformation:
\bea
\tkm{3}{n} &=& \km{3}{n} - \frac{k}{2} w \delta_n, \\
\tkm{+}{n} &=& \km{+}{n+w}, \\
\tkm{-}{n} &=& \km{-}{n-w}.
\eea
Then, the Virasoro generators $\Lm{m}$ are related to $\tLm{m}$ by
\be
\Lm{m} = \tLm{m} -w \tkm{3}{m}-\frac{k}{4} w^2 \delta_m.
\ee
This transformation is known as the spectral flow. References~\cite{Henningson:1991jc,Maldacena:2000hw} proposed to include the representations transformed by the spectral flow. Denote the resulting representations by ${\hat{\cal D}}^{\pm,w}_{\tj}$ and ${\hat{\cal C}}^{\alpha,w}_{\tj}$, where $\tj$ labels the spin before the flow (Similarly, $\tilde{m}$ and $\tilde{N}$). The on-shell condition becomes
\be
\Lm{0} = -\frac{\tj(\tj-1)}{k-2} - w \tilde{m} - \frac{k}{4} w^2 + \tilde{N} + \hKp =1.
\ee
In general, $\Lm{0}$ is not bounded below for the flowed representations. However, ${\hat{\cal D}}^{\pm,w=\mp 1}_{\tj} = {\hat{\cal D}}^{\mp}_{m-\tj}$. Thus, in order for the discrete representations with $j, \tj<1/2$ not to appear, the spin $j$ has to satisfy the bound
\be
\frac{1}{2}<j<\frac{k-1}{2}.
\label{eq:bound2}
\ee
Note that this bound is stronger than the bound \eq{bound1}, which is needed to show the no-ghost theorem. When ${\hat{\cal D}}^{\pm,w}_{\tj}$ saturates the lower bound of Eq.~\eq{bound2}, ${\hat{\cal C}}^{\alpha,w}_{\tj}$ appears; likewise when ${\hat{\cal D}}^{\pm,w}_{\tj}$ saturates the lower bound of Eq.~\eq{bound2}, ${\hat{\cal C}}^{\alpha,w+1}_{\tj}$ appears \cite{Maldacena:2000hw}. The representations ${\hat{\cal C}}^{\alpha,w}_{\tj}$ correspond to long strings \cite{Maldacena:1998uz,Seiberg:1999xz}.

Then, Ref.~\cite{Maldacena:2000hw} shows the no-ghost theorem for the spectral flowed representations with the bound \eq{bound1}.

\section{The outline of the proof}\label{sec:outline}

Now, we turn into the discussion of the no-ghost theorem in the BRST quantization. The full proof of the no-ghost theorem is rather involved and we will use our previous proofs. Thus, we briefly describe general structure of our proof here. The terminology appeared below is explained later. In general, the proof of the no-ghost theorem consists of 2 steps in the BRST quantization (\tabl{outline}).

\begin{table}
\begin{center}
\begin{tabular}{lc}
Step~0:& Matter Hilbert space via Verma modules \\
&$\downarrow$ \\
Step~1:& The vanishing theorem using filtration \\
&(reason why $d \geq 2$ in standard proofs) \\
&$\downarrow$ \\
Step~2:& The no-ghost theorem
\end{tabular}
\caption{The outline of the proof.}
\label{table:outline}
\end{center}
\end{table}

\begin{itemize}

\item Step~1:
The first is to show the {\it vanishing theorem}. The vanishing theorem states that the $\hQ$-cohomology is trivial except at the zero ghost number. This is done by choosing an appropriate {\it filtration} for your BRST operator $\hQ$. A filtration allows us to use a simplified BRST operator $Q_{0}$ and we can first study the cohomology of $ Q_{0} $. If the $ Q_{0} $-cohomology is trivial, so is the $\hQ$-cohomology (Lemma~3.1 of Ref.~\cite{Asano:2000fp}); this is the reason why the filtration is so useful. However, the particular filtration used in standard proofs is also part of the reason why $d \geq 2$ in those proofs. 

\item Step~2:
The second is to compute and compare the {\it index} and the {\it signature} of the cohomology group explicitly. If the index is equal to the signature, the no-ghost theorem holds provided the vanishing theorem is valid. 

\end{itemize}

\noindent
Step~1 and 2 themselves consist of several steps, which are explained in \sect{vanishing} and \ref{sec:no-ghost:ads3}, respectively.

However, in our approach it is useful to have an additional step:

\begin{itemize}
\item Step~0:
Write the matter Hilbert space in terms of products of two Verma modules, one for a nonunitary $c=1$ CFT and the other for a unitary CFT $K$.
\end{itemize}

\noindent Using this form of the Hilbert space, we have shown Step~1 or the vanishing theorem in our previous papers. (The $d=1$ case is a particular example.) So, the vanishing theorem for $AdS_3$ follows immediately if one can show Step~0 for $AdS_3$. This will be our main focus in the next section.  Then, we will show Step~2 in \sect{no-ghost:ads3}.


Incidentally, to establish Step~1 in our previous works, we do not use the standard filtration, but use the filtration introduced by Frenkel, Garland, and Zuckerman \cite{Frenkel:1986dg}. In our approach, 
the proof refers only to the matter Virasoro generators themselves. This is convenient when one discusses general spacetime backgrounds such as $AdS_3$.

\section{The vanishing theorem for string theory}\label{sec:vanishing}

We start with our vanishing theorem \cite{Asano:2000fp,Asano:2003jn}.  
Let ${\cal V}(c, h)$ be a Verma module with highest weight $h$ and central charge~$c$. Then, 
\begin{theorem}[The vanishing theorem for string theory]
The $\hQ$-cohomology can be non-zero only at zero ghost number 
if the matter part of the Hilbert space $ {\cal H} $ can be decomposed as a sum of the following two Verma modules:
\begin{equation}
{{\cal H}_{\hX,\hK} = \cal V}(\cX=1, \hX<0) \otimes {\cal V}(\cK=25, \hK>0).
\label{eq:hilbert}
\end{equation}
(Or more generally, the $\cX=1$ part is written as nondegenerate Verma modules.)
\label{thm:vanishing}
\end{theorem}
\noindent 
Here, the restriction on the weights $\hX$ and $\hK$ comes in order for the Verma modules to form the basis of the $\cX=1$ nonunitary CFT and the $\cK=25$ unitary CFT, respectively.
Thus, the vanishing theorem follows immediately once one establishes that the theory in question has the Hilbert space in the form of Eq.~\eq{hilbert}, which is Step~0 of \sect{outline}. 

The proof of this vanishing theorem consists of three steps (\tabl{vanishing}):

\begin{itemize}
\item[$\circ$] Step~1.1:
Apply our filtration \'{a} la Frenkel, Garland, and Zuckerman (Ref.\cite{Frenkel:1986dg}; Eq.~(35) of Ref.~\cite{Asano:2000fp}). With FGZ's filtration, the simplified BRST operator $Q_0$ can be further decomposed as a sum of two differentials, $d'$ and $d''$. This decomposition is crucial for the proof; it reduces the problem to the $d'$-cohomology, and $d'$ acts only on the $\cX=1$ part and the $b$ ghost part.
This is the reason why the proof does not require $d \geq 2$.
\item[$\circ$] Step~1.2:
If the $d'$-cohomology is trivial, so is the $Q_0$-cohomology. This follows from a K\"{u}nneth formula. Then, the $\hQ$-cohomology is trivial as well from the general property of filtration.
\item[$\circ$] Step~1.3:
Now, the problem is reduced to the $d'$-cohomology. Show the vanishing theorem for the $d'$-cohomology. 
%
\end{itemize}
\noindent
See Refs.~\cite{Asano:2000fp,Asano:2003jn} for the actual proof.

\begin{table}
\begin{center}
\begin{tabular}{c}
Filtration (Step~1.1) \\
$\downarrow$ \\
$Q^{\rm (FGZ)}_{0} = d'+d''$ \\
effectively reduces the problem to the $\cX=1$ part \\ 
$\downarrow$ \\
The Vanishing Theorem for the $d'$-cohomology (Step~1.3) \\
$\downarrow$ \\
The Vanishing Theorem for the $Q^{\rm (FGZ)}_{0}$-cohomology (Step~1.2) \\
$\downarrow$ \\
The Vanishing Theorem for the $\hQ$-cohomology (from the general property of filtration) \\
\end{tabular}
\caption{The outline of the proof of the vanishing theorem.}
\label{table:vanishing}
\end{center}
\end{table}

In addition to the vanishing theorem, we have actually established a stronger statement in the paper:
\begin{theorem}[FP-ghost decoupling theorem]
Physical states do not contain Fadeev-Popov ghosts if $ {\cal H} $ is decomposed as in Eq.~\eq{hilbert}.
\label{thm:decoupling}
\end{theorem}
\noindent
Although the theorem itself is not necessary to establish the no-ghost theorem, it is useful to establish, {\it e.g.}, the BRST-OCQ equivalence (\sect{equiv}). 


We discuss the examples where our vanishing theorem can be applied. Again, one merely has to check that a theory has the Hilbert space in the form of Eq.~\eq{hilbert}.

\subsection{$d=1$ case}\label{sec:d=1}

For $\IR^{1,d-1} \times K (d=1)$, choose the $\cX=1$ CFT as a CFT generated by the timelike oscillators $\alpha^{0}_{m}$. The isomorphism
\begin{equation}
\mbox{Fock}(\alpha^{0}_{-m}, ;k^0) \cong {\cal V}(1, \hX) 
\cond{if $\hX=\alp k^2<0$}
\label{eq:verma}
\end{equation}
can be easily shown using the Kac determinant for $\cX=1$ (Eq.~(45) in Ref.~\cite{Asano:2000fp}). The CFT $K$ is a unitary CFT, so $\hK>0$.

\subsection{Flat spacetime or $d \geq 2$ cases}

Again, choose the $\cX=1$ CFT as a CFT generated by the timelike oscillators $\alpha^{0}_{m}$. The $\cK=25$ CFT is a product of the CFT generated by $\alpha^{i}_{m}$ and a unitary CFT $K'$. When $(k^i)^2>0$, the $\cK=25$ CFT can be written in terms of Verma modules with $\hK>0$ \cite{Frenkel:1986dg}, so reduces to the $d=1$ case.

\subsection{$AdS_3$}

The Hilbert space of the $SL(2,\IR)$ WZW model can also be decomposed as in Eq.~\eq{hilbert} under the restriction on the spin \eq{bound1}. This construction of the Hilbert space has been done in Refs.~\cite{Hwang:1990aq,Hwang:1991an,Evans:1998qu,Maldacena:2000hw}; it was discussed in order to prove the no-ghost theorem for $AdS_3$ in the context of the old covariant quantization. 

Choose the $\cX=1$ CFT as a CFT generated by $J^{3}_{n}$, which corresponds to the timelike $U(1)$: 
\be
\Lt{m}=-\frac{1}{k} \sum_n \nor \km{3}{m-n} \km{3}{n} \nor.
\ee
Also, define 
\be
\LK{m} = \Lm{m} -\Lt{m}.
\ee 
By construction, $\LK{m}$ commute with $\km{3}{n}$ and thus with $\Lt{n}$. Thus, as far as Virasoro generators are concerned, the $SL(2,\IR)$ WZW model can be decomposed as
\be
U(1) \times SL(2,\IR)/U(1)
\label{eq:decomposition}
\ee
with the Hilbert space consisting of all states of the form
\begin{equation}
\Lt{-m_1} \Lt{-m_2} \ldots \Lt{-m_M} 
\LK{-n_1} \LK{-n_2} \ldots \LK{-n_N} \ket{\hX}\ket{\hK},
\label{eq:basis}
\end{equation}
where $ m_{1} \leq m_{2} \leq \cdots \leq m_{M} $ and $ n_{1} \leq n_{2} \leq \cdots \leq n_{N} $. 
Thus, we regard the coset part as $K$. When there is an internal CFT $K'$, we include it as a part of $K$. 
Note that even though we write $U(1) \times SL(2,\IR)/U(1)$, it does not mean that the $U(1)$ part is completely separated from the coset part as a direct product; the $\km{3}{0}$-value for the Virasoro primaries must be common in both.

Since such a construction of the Hilbert space is fairly standard \cite{Goddard:1972iy}, here we merely check that the weights $\hX$ and $\hK$ are as in Eq.~\eq{hilbert}. We discuss the unflowed and flowed representations separately.

{\it (i) Unflowed representations:} The $\cX=1$ CFT has 
\be
\hX=-\frac{m^2}{k} \leq 0.
\ee
Thus, the CFT can be written in terms of the Verma modules ${\cal V}(\cX=1, \hX<0)$ except $m=0$ states. However, there is no on-shell state with $m=0$ except a few states, which have zero ghost number and have positive norm, so they do not affect the vanishing theorem and the no-ghost theorem \cite{Evans:1998qu,Maldacena:2000hw}. 
The coset can also be written in terms of the Verma modules ${\cal V}(\cK=25, \hK>0)$. For representations ${\hat{\cal D}}^+_j$, this is possible if the spin $j$ satisfies the bound \eq{bound1} \cite{Evans:1998qu,Maldacena:2000hw}. If the Virasoro primary $\ket{\hK}$ is at grade $M$ of ${\hat{\cal D}}^+_j$,
\bea
\hK &=& -\frac{j(j-1)}{k-2}+M+\frac{m^2}{k}+\hKp 
\nonumber \\
&=& \frac{2j(k/2-j)}{k(k-2)}+\frac{2M}{k}(\frac{k}{2}-j)+\frac{2j}{k}(-j+m+M)+\frac{1}{k}(j-m)^2+\hKp
>0
\eea
since $k>0$, $j>0$, $k/2-j>0$, $-j+m+M \geq 0$, and $\hKp \geq 0$. Here, the bound $-j+m+M \geq 0$ comes from the $SL(2,\IR)$ Clebsh-Gordon decomposition: the spin at grade $M$ is from $j-M$ to $j+M$. Note that $m$ here is the total $\km{3}{0}$-value, not the $\km{3}{0}$-value for ${\cal D}^+_j$. The weight $\hK$ is non-negative because the coset is unitary under the bound \cite{Dixon:1989cg}.

{\it (ii) Flowed representations:} The discussion for the flowed representations ${\hat{\cal D}}^{+,w}_{\tj}$ and ${\hat{\cal C}}^{\alpha,w}_{\tj}$ is similar. For the flowed representations, 
\bea
\Lt{0} &=& \tLt{0} -w \tkm{3}{0}-\frac{k}{4} w^2, \\
\LK{0} &=& \tLm{0} - \tLt{0} = \tilde{L}_0^{\rm K}.
\eea 
The $\cX=1$ CFT can be written in terms of the Verma modules since
\be
\hX = - \frac{m^2}{k} = -\frac{(\tilde{m}+\frac{k}{2}w)^2}{k} \leq 0.
\ee
Again the $m=0$ state may be problematic, but one can check that there is no $m=0$ on-shell state except the ground state \cite{Maldacena:2000hw}. 
The coset can also be written in terms of the Verma modules since 
\be
\hK = -\frac{\tilde{j}(\tilde{j}-1)}{k-2}+\tilde{M}+\frac{\tilde{m}^2}{k}+\hKp>0
\ee
for ${\hat{\cal D}}^{+,w}_{\tj}$ and
\be
\hK = \frac{\tilde{s}^2+\frac{1}{4}}{k-2}+\tilde{M}+\frac{\tilde{m}^2}{k}+\hKp>0
\ee
for ${\hat{\cal C}}^{\alpha,w}_{\tj}$.


\section{The no-ghost theorem for $AdS_3$}\label{sec:no-ghost:ads3}

\begin{table}
\begin{center}
\[
\begin{array}{ccc}
\mbox{tr}_{\obs} \, q^{\Lint} C 
	&\stackrel{\mbox{\footnotesize No-Ghost Theorem}}{=}& 
	\mbox{tr}_{\obs} \, q^{\Lint} \\
&& \\
	\mbox{\footnotesize Step~2.2} \updownarrow 
	&& \updownarrow \mbox{\footnotesize Step~2.1} \\
&& \\
\mbox{tr}_{\hat{\cal H}} \, q^{\Lint} C 
	&\stackrel{\mbox{\footnotesize Step~2.3}}{\leftrightarrow}& 
	\mbox{tr}_{\hat{\cal H}} \, q^{\Lint} (-)^{\hN}
\end{array}
\]
\caption{Strategy to prove the no-ghost theorem. The traces ``$\mbox{tr}_{\obs}$" and ``$\mbox{tr}_{\hat{\cal H}}$" are taken over the observable Hilbert space and the Hilbert space $\hat{\cal H}$, respectively. The operator $C$ gives eigenvalues $C_{a}$ and $\Lint$ counts the total grades.  The no-ghost theorem reduces a calculations of weighted characters modulo the on-shell condition. }
\label{table:noghost}
\end{center}
\end{table}

In order to show the no-ghost theorem, the notion of {\it signature} is useful. For a vector space $V$ with an inner product, we can choose a basis $e_{a}$ such that
$\norm{e_{a}}{e_{b}} = \delta_{ab} C_{a},$
where $C_{a}\in \{0, \pm 1 \}$. Then, the signature of $V$ is defined as
$\sig{V} = \sum_{a} C_{a}$. If $ \sig{V} = \dims{V} $, all the $C_{a}$ are 1, so $V$ has positive definite norm. Then, the statement of the no-ghost theorem is equivalent to 
\be
\mbox{tr}_{\obs} \, q^{-L_{0}^{\rm mass}} C = \mbox{tr}_{\obs} \, q^{-L_{0}^{\rm mass}},
\label{eq:noghost1}
\ee
where $q=e^{2\pi i \tau}$, the trace ``$\mbox{tr}_{\obs}$" is taken over the observable Hilbert space, and the operator $C$ gives eigenvalues $C_{a}$. Here, $L_{0}^{\rm mass}$ is the ``mass" term in $L_{0}$ ,{\it i.e.}, the $L_{0}$-eigenvalue for the ground state [$\alp k^2$ for the flat spacetime and $c_2/(k-2)$ for $AdS_3$]. Using the on-shell condition
\be
L_{0} \ket{\phi} = (L_{0}^{\rm mass}+\Lint) \ket{\phi} = 0
\label{eq:on-shell2}
\ee
($\Lint$ counts the total grades.), Eq.~\eq{noghost1} becomes
\be
\mbox{tr}_{\obs} \, q^{\Lint} C = \mbox{tr}_{\obs} \, q^{\Lint}.
\label{eq:noghost2}
\ee

Equation~\eq{noghost2} can be shown with 3 steps (\tabl{noghost}). This has been done repeatedly for $d \geq 2$ in Refs.~\cite{Frenkel:1986dg,Lian:cy,Figueroa-O'Farrill:1989mw,Figueroa-O'Farrill:1989hu}. 
Step~2.1 and 2.2 make use of the BRST quartet mechanism and Step~2.1 uses {\it the vanishing theorem} as well. Step~2.3 can be done by explicitly calculating the both sides. Once one establishes the vanishing theorem, the only nontrivial step is Step~2.3, {\it i.e.}, to show
\begin{equation}
\mbox{tr}_{\hat{\cal H}} \, q^{\Lint} C
        = \mbox{tr}_{\hat{\cal H}} \, q^{\Lint} (-)^{\hN}.
\label{eq:noghost3}
\end{equation}
Note that the trace weighted by $(-)^{\hN}$ is an {\it index}. 

For the $AdS_3$ case, we also make use of closely related quantities: 
\be
\mbox{tr} \, q^{\Lint} C, \qquad
\mbox{tr} \, q^{\Lint} (-)^{\hN},
\label{eq:characters}
\ee
where the trace ``$\mbox{tr}$" is taken over the Hilbert space ${\cal H}$. These are essentially weighted characters of the Virasoro algebra. The difference between $\hat{\cal H}$ and ${\cal H}$ is that the latter does not impose the on-shell condition. 


\subsection{Flat spacetime or $d \geq 2$ cases}

In standard proofs, one explicitly computes the timelike, longitudinal oscillator parts and the $(b,c)$ ghosts part. Then, one gets
\be
\mbox{tr}_{\hat{\cal H}} \, q^{\Lint} C
= \mbox{tr}_{\hat{\cal H}} \, q^{\Lint} (-)^{\hN}
= \mbox{tr}_{{\cal H}_{K}} \, q^{\LK{0}-1}.
\ee
Here, $K$ is the product of a transverse CFT and a unitary CFT $K'$. This also shows the equivalence with the light-cone spectra.

\subsection{$d \geq 1$ cases}

In our previous papers, we pointed out that the standard proof is also valid for $d=1$ by computing the $\cX=1$ part and the $(b,c)$ ghosts part if the vanishing theorem is valid for $d=1$. Then, one gets
\bea
\mbox{tr}_{\hat{\cal H}} \, q^{\Lint} C
= \mbox{tr}_{\hat{\cal H}} \, q^{\Lint} (-)^{\hN}
&=& q^{-1} \prod_{m} (1-q^{m}) \, \mbox{tr}_{{\cal H}_{K}} \, q^{\LK{0}}
\nonumber \\
&=& \eta(\tau) \, \mbox{tr}_{{\cal H}_{K}} \, q^{\LK{0}-\frac{\cK}{24}},
\eea
where $\eta(\tau)$ is the Dedekind eta function.

\subsection{$AdS_3$}\label{sec:noghost-ads3}

In general, the computation of the index and signature may not be easy and one needs to take an orthonormal basis for the signature. However, for the $AdS_3$ case, the $\cX=1$ part is written in terms of a free CFT and the coset is unitary just like $K$. Then, the situation is the same as the $d=1$ case and the computation is essentially the same. Thus, we mainly discuss some minor complications specific to $AdS_3$.

%
%
The Hilbert space consists of a sum of Verma modules of the form \eq{basis}. We consider the index and the signature for each Verma module with given $\hX$ and $\hK$. Only positive $\hK$ is considered since the no-ghost theorem is expected to hold only for this case. The following basis is taken for Step~2.3:
\begin{itemize}
\item $U(1)$ part: $\km{3}{-n}$
\item The coset part: the basis \eq{basis}, which is further diagonalized (although we do not have to compute it explicitly.)
\end{itemize}
\noindent
It is important here that we take the basis \eq{basis}, not $\km{a}{-n}$ since $\km{3}{-n}$ do not commute with $\km{a}{-n}$. 

First, let us consider the weighted $SL(2,\IR)$ characters \eq{characters}. For all the representations we consider, 
\be
\tr q^{\Lint} (-)^{\hN} 
= \eta(\tau) \, \mbox{tr}_{{\cal H}_{K}(\hK>0)} \, q^{\LK{0}-\frac{\cK}{24}}. 
\label{eq:character1}
\ee
Here, $\mbox{tr}_{{\cal H}_{K}(\hK>0)}$ is taken over the Hilbert space ${\cal H}_{K}$ with a given $\hK>0$. For flowed representations, one had better use $\tLm{0}$ instead of $\Lm{0}$ for the exponent of $q$; otherwise, powers of $q$ are not directly related to the mass. However, it is sufficient to use $\Lm{0}$ for our purpose. 

The above weighted character \eq{character1} takes the trace over ${\cal H}$, whereas the index \eq{noghost3} takes the trace over $\hat{\cal H}$. Thus, the weighted character does not really represent the dimension of physical space. 
The on-shell condition needs to be imposed in order to get the dimension from Eq.~\eq{character1}. This gives rise to two issues. First of all, given a series of representations for $AdS_3$, the on-shell condition may not be satisfied for all grades. For the flat spacetime, the on-shell condition is always satisfied by choosing momentum $k^{\mu}$ appropriately. However, in general this may not be the case, and this is not the case for $AdS_3$. For example, a continuous representation ${\hat{\cal C}}^\alpha_j$ typically appears only at the ground state $N=0$ when $k>2$.


Second, for the discrete series ${\hat{\cal D}}^+_j$ and ${\hat{\cal D}}^{+,w}_{\tj}$, one must impose the bound \eq{bound1}.
The bound in turn implies that the grade is bounded above \eq{bound_for_grade} from the on-shell condition. 
Thus, the weighted character can be interpreted as the dimension at a particular grade if the grade is consistent with the on-shell condition and the bound. In other words, even though we sum over all the grades in the character, the full spectrum implied from the character does not appear as physical states. However, these are minor points since weighted characters for $AdS_3$ are the same even before we impose the on-shell condition [See \eq{character_equality}].

Now, for the character weighted by $C$, 
\be
\tr q^{\Lint} C 
= \eta(\tau) \, \mbox{tr}_{{\cal H}_{K}(\hK>0)} \, q^{\LK{0}-\frac{\cK}{24}},
\label{eq:character2}
\ee
where we used the unitarity of the $SL(2,\IR)/U(1)$ coset for $\hK>0$ \cite{Dixon:1989cg}. 
Comparing Eqs.~\eq{character1} and \eq{character2}, we get
\be
\mbox{tr} \, q^{\Lint} C = \mbox{tr} \, q^{\Lint} (-)^{\hN}.
\label{eq:character_equality}
\ee
Then, taking the on-shell condition into account, we established
\be
\mbox{tr}_{\hat{\cal H}} \, q^{\Lint} C
        = \mbox{tr}_{\hat{\cal H}} \, q^{\Lint} (-)^{\hN}.
\ee


Note that the timelike direction cancels with one of the FP-ghost contribution in Eq.~\eq{character1}, but the character still has a factor of $\eta(\tau)$. Consequently, the dimension of the physical Hilbert space is smaller than the Hilbert space of $K$ (by the dimension of a nondegenerate $c=1$ Verma module). This is because null states arise by tensoring the $\cX=1$ CFT and $K$. In the OCQ language, this means that the physical spectrum includes neither timelike oscillators nor OCQ-null states and consists only of DDF states (\append{appC}).

One would recognize that the character \eq{character1} is in fact part of the full $SL(2,\IR)$ characters (times the FP-ghost contribution); {\it e.g.}, see Appendix~B of Ref.~\cite{Maldacena:2000hw}. One way to obtain the modular invariant partition function from the character is as follows \cite{Maldacena:2000hw}%
\footnote{Also, see Refs.~\cite{Petropoulos:1989fc,Petropoulos:1999nc,Henningson:1991jc} and Refs.~\cite{Griffin:1990fg}-\cite{Maldacena:2000kv}.}: 
First, sum over the $\km{3}{0}$-eigenvalue with weight $z^{\km{3}{0}}$. The weight is necessary to avoid the divergence which arises due to the infinite degeneracy of zero mode representations. 
Next, take the ``diagonal combination" of character with the anti-holomorphic part. Then, sum over all the representations. Finally, take into account the chiral anomaly. 
The partition function obtained in this way is identical to the one computed for $SL(2,C)/SU(2)$ model \cite{Gawedzki:1991yu}, which is expected to be related to the $SL(2,\IR)$ model by some Euclidean rotation.

\section{The no-ghost theorem for string theory}\label{sec:no-ghost}

We saw that the vanishing theorem is valid not only for $d=1$, but also valid as long as the matter part of the Hilbert space ${\cal H}$ is written as a sum of two Verma modules: 
\begin{equation}
{\cal H}_{\hX,\hK} = 
{\cal V}(\cX=1, \hX<0) \otimes {\cal V}(\cK=25, \hK>0).
\label{eq:hilbert2}
\end{equation}
Moreover, the no-ghost theorem is valid for such a Hilbert space as well. To see this, note that the isomorphism \eq{verma} works in both directions. In one direction, we can use the isomorphism to write the $\cX=1$ Fock space in terms of a Verma module, but given a Hilbert space ${\cal H}$ in terms of Verma modules as in Eq.~\eq{hilbert2}, one can take the basis of a {\it nonunitary} free boson $\alpha^{0}_{-m}$ for the $\cX=1$ part. Then, the proof of the no-ghost theorem is essentially the same as the $d=1$ or $AdS_3$ case. 

There is one difference however. The computation of the signature assumes that Virasoro primaries $\ket{\hX}$ have positive-definite norm. This is trivial for the $d=1$ case; the Virasoro primaries in this case are just $\ket{k^0}$. However, this point is nontrivial in general and one has to check this point separately. This is similar to the requirement in the OCQ that DDF states have positive-definite norm. Thus, we arrives at the following theorem:
%
%
\begin{theorem}[The no-ghost theorem for string theory]
$\hat{\cal H}_{\obs}$ is a positive definite space (i) if the matter part of the Hilbert space ${\cal H}$ is decomposed as in Eq.~\eq{hilbert},
and (ii) if the Virasoro primaries $\ket{\hX}$ used to construct the $\cX=1$ Verma module have positive-definite norm.
\label{thm:no-ghost}
\end{theorem}
\noindent
Let us expand condition~(ii) more (in the OCQ language). If there were a Virasoro primary in the $\cX=1$ part other than a ground state, it would imply that the primary is the Virasoro primary under the full Virasoro algebra. The only case that this is not a OCQ-physical state is when this is a null state, but we assume that there is no such null state in condition~(i). Then, the timelike direction is not decoupled and there is a physical state with the timelike polarization. This does not necessarily mean that the theory is problematic. However, if the norm of such a Virasoro primary is not positive-definite, it would imply the violation of the no-ghost theorem. 

It is instructive to see backgrounds {\it with ghosts} and see how these backgrounds violate the assumptions of our no-ghost theorem. 
The Lorentzian orbifold $\IR^{1,\tilde{d}}/{\bf Z}_2$ is an example \cite{Balasubramanian:2002ry}. In certain cases, this example contains ghosts, so our no-ghost theorem should not apply. This example does not satisfy our no-ghost theorem partly because it violates condition~(ii) of \thm{no-ghost}. 
They consider the spacetime under the following action (For simplicity, consider the bosonic string.):
\be
X^a \rightarrow -X^a \quad (a=0,\ldots,\tilde{d}), \qquad
X^i \rightarrow X^i \quad (i=\tilde{d}+1, \ldots, 25)
\ee
They found a ghost when $\tilde{d} \leq 7$ and we focus on these cases. In the twisted sector, $X^a$ are antiperiodic; so there is no momentum $k^a$ and the zero-point energy is shifted. Thus,
\bea
\hX &=& \frac{1}{16} + M >0, \\
\hK &=& \alp (k^i)^2 +\frac{\tilde{d}}{16} + M' >0,
\eea
where Virasoro primaries $\ket{\hX}, \ket{\hK}$ are assumed to be at grade $M, M'$, respectively. Thus, the matter Hilbert space is not decomposed as in Eq.~\eq{hilbert}, but one can still form the $\cX=1$ Hilbert space by Verma modules since the resulting Kac determinant has no zeros. In fact, the character for the $c=1$ twisted sector is decomposed in terms of a free boson character $\eta(\tau)^{-1}$:
\bea
\chi_{c=1,h=\frac{1}{16}}(q) 
&=& q^{\frac{1}{48}}\prod_{m=1}^{\infty} (1-q^{m-\frac{1}{2}})^{-1} 
\nonumber \\
&=& q^{\frac{1}{16}} \eta(\tau)^{-1} 
	\sum_{n=1}^{\infty} q^{\frac{n(n-1)}{4}} 
\nonumber \\
&=& q^{\frac{1}{16}} \eta(\tau)^{-1} 
	(1 + q^{\frac{1}{2}} + q^{\frac{3}{2}} + \cdots).
	\label{eq:z_2character}
\eea

However, the $\cX=1$ Hilbert space does not satisfy the condition~(ii) of \thm{no-ghost}. For example, the Hilbert space contains a Verma module constructed from a Virasoro primary $\alpha^0_{-1/2} \ket{0}$ and this gives rise to negative-norm states. [The primary corresponds to the $q^{\frac{1}{2}}$ term in Eq.~\eq{z_2character}.] 

Also, one can write each $\cX=1$ Verma module in terms of a free boson basis, but the resulting free boson basis is unitary since the $\cX=1$ part has positive $\hX$. (The nonunitarity of the model comes from the nonunitarity of the Virasoro primaries.) The weighted characters for this background \eq{characters} are not the same due to these two reasons.

Fortunately, for the Lorentzian orbifold, both $\hX$ and $\hK$ are positive so that there is only finite number of on-shell states; physical spectrum can be computed explicitly and there is no ghost when $\tilde{d} > 7$. 

\section{BRST-OCQ equivalence}\label{sec:equiv}

For string theory on curved backgrounds, one often uses the OCQ. The purpose of this section is to establish the connection between the BRST quantization discussed here and the OCQ. The BRST-OCQ equivalence is well-established for the flat case. When the background is curved, the equivalence may not hold although it is certainly likely. Here, we prove the BRST-OCQ equivalence for the backgrounds where \thm{no-ghost} hold. We sketch the standard textbook proof \cite{Polchinski:rq} emphasizing the necessary assumptions and the differences from the standard case. As is clear from below, the proof requires the following assumptions:
\begin{enumerate} 
\item The BRST version of the no-ghost theorem
\item For each physical state, there is an equivalent class where the FP-ghost sector is in ghost vacuum $\ket{\downarrow}$.
\end{enumerate}

What we have to show is that there is a map which maps the OCQ equivalence classes to the BRST equivalence classes, and the map is both one-to-one and onto. Let $\ket{\psi}$ be a state in the matter Hilbert space of the OCQ. Associate a state 
\be
\ket{\psi, \downarrow}
\label{eq:ghost_ground}
\ee
from the Hilbert space in the BRST quantization. Then, the proof consists of 4 steps. 

\begin{itemize}
\item Step~1:
{\it OCQ physical state $\rightarrow$ BRST closed state} \\
Show that each OCQ physical state maps to a BRST-closed state. This follows from
\be
Q \ket{\psi, \downarrow} = \sum_{n=0}^{\infty} c_{-n} (\Lm{n} - \delta_{n}) \ket{\psi, \downarrow}.
\ee

\item Step~2:
{\it Equivalent OCQ states $\rightarrow$ Equivalent BRST states} \\
Show that if $\ket{\psi}$ and $\ket{\psi'}$ are equivalent OCQ physical states, they map into the same BRST class:
\be
\ket{\psi,\downarrow}-\ket{\psi',\downarrow} = Q \ket{\chi}. 
\label{eq:equiv}
\ee
The state is a BRST-closed state from Step~1 and has a zero norm since $\ket{\psi}-\ket{\psi'}$ is OCQ-null. From the no-ghost theorem (BRST), a zero-norm closed state must be exact. So, this step uses the no-ghost theorem.

\item Step~3:
{\it The map is one-to-one, {\it i.e.}, Equivalent OCQ states $\leftarrow$ Equivalent BRST states.}\\
The converse to Step~2. Show that $\ket{\psi}-\ket{\psi'}$ is OCQ null if Eq.~\eq{equiv} holds. One expands $\ket{\chi}$ in terms of excitations over the ghost ground state, substitute it into Eq.~\eq{equiv}, and compare the both sides. 

\item Step~4:
{\it The map is onto.} \\
Show that every BRST class contains at least one state of the form \eq{ghost_ground}. In our case, this follows from the FP-ghost decoupling theorem (\thm{decoupling}).
\end{itemize}

We have therefore established the no-ghost theorem in the OCQ, for the backgrounds which satisfy \thm{no-ghost}, in particular for $AdS_3$. In fact, one can actually show the no-ghost theorem in the OCQ directly under the same assumption as the BRST quantization \cite{Hwang:1990aq,Hwang:1991an,Evans:1998qu,Maldacena:2000hw}. Specifically, as discussed in \append{appC}, the proof requires that 
\begin{itemize}
\item[$\circ$] The Hilbert space is written as in Eq.~\eq{hilbert}.
\end{itemize}
This assumption is the same as ours.

\section*{Note Added}

While this paper is in preparation, we received a preprint by Pakman \cite{Pakman:2003kh} which has some overlap with the present paper. The no-ghost theorem on a generic spacetime has been also discussed in Ref.~\cite{Hwang:1998tr}.


\acknowledgments

We would like to thank Vijay Balasubramanian, Mirjam Cvetic, Yang-Hui He, Asad Naqvi, Burt Ovrut, and Joe Polchinski. We would especially like to thank Mitsuhiro Kato for continuous help throughout our projects and Yuji Satoh for useful discussions and for various comments on the draft. M.N.\ would like to thank theoretical high energy physics group at Univ.\ of Pennsylvania for the kind hospitality where this work was carried out. The research of M.A. was supported in part by the fellowship from {\it Soryushi Shogakukai}. The research of M.N.\ was supported in part by the Grant-in-Aid for Scientific Research (13740167) and the Oversea Research Fellowship from the Ministry of Education, Culture, Sports, Science and Technology, Japan.

\appendix

\section{Notations and conventions}\label{sec:appA}

We use the notations and conventions of Refs.~\cite{Asano:2000fp,Asano:2003jn}. 
We assume that the total $L_{m}$ of the theory is given by 
\be
L_{m} = \underbrace{\Lt{m} + \LK{m}}_{\textstyle \Lm{m}} + \Lg{m}. 
\ee
We assume that $K$ is a unitary CFT and all states in $K$ lie in highest weight representations. 

The ghost number operator $\hN$ is normalized so that $\hN \ket{\downarrow}=0$ since the ghost zero modes will not matter to our discussion.

We will call the total Hilbert space $\Htotal$, but it is useful to define the following subspaces:
\begin{eqnarray}
{\cal H} &=& \{ \phi \in \Htotal: b_{0} \phi = 0 \},\\
\hat{\cal H} &=& \{ \phi \in \Htotal: b_{0} \phi = L_0 \phi = 0 \}.
\end{eqnarray}
The physical state conditions are
\be
Q \ket{\mbox{phys}} = 0, \quad b_{0} \ket{\mbox{phys}} = 0, \quad L_{0} \ket{\mbox{phys}} = 0.
\ee
Thus, we will consider the cohomology on $\hat{\cal H}$; $Q$ takes $\hat{\cal H}$ into itself. The Hilbert space $\hat{\cal H}$ is classified according to mass eigenvalues. The on-shell condition is written as
\be
L_{0} \ket{\phi} = (L_{0}^{\rm mass}+\Lint) \ket{\phi} = 0,
\ee
where $L_{0}^{\rm mass}$ is the ``mass" term in $L_{0}$ ,{\it i.e.}, the $L_{0}$-eigenvalue for the ground state [$\alp k^2$ for the flat spacetime and $c_2/(k-2)$ for $AdS_3$], and $\Lint$ counts the total grade.

Decompose the BRST operator $Q$ in terms of ghost zero modes:
\be
Q = \hQ
+ (\mbox{terms in $Q$ with ghost zero modes}).
        \label{eq:hatq}
\ee
Then, for a state $\ket{\phi} \in \hat{\cal H}$,
\begin{equation}
Q \ket{\phi} = \hQ \ket{\phi}.
        \label{eq:relative}
\end{equation}
Therefore, the physical state condition reduces to
\begin{equation}
\hQ \ket{\phi} = 0.
\end{equation}
Also, $\hQ^2=0$ on $\hat{\cal H}$ from Eq.~\eq{relative}. Thus, $\hQ$ defines a BRST complex, which is called the {\it relative} BRST complex. So, we can define $\Hc, \He \subset \hat{\cal H}$ by
\begin{equation}
\hQ \Hc = 0, \qquad
\He = \hQ \hat{\cal H},
\end{equation}
and define the relative BRST cohomology of $Q$ by
\begin{equation}
\hat{\cal H}_{\obs}=\Hc/\He.
\end{equation}

\section{The no-ghost theorem in the OCQ: $d=1$ and $AdS_3$ cases}\label{sec:appC}

Here, we sketch the proof for $d=1$ and $AdS_3$ in the old covariant quantization. This has been done in Refs.~\cite{Hwang:1990aq,Hwang:1991an,Evans:1998qu,Maldacena:2000hw}. The original proof of Goddard-Thorn \cite{Goddard:1972iy} itself does not hold for these cases, but one can extend to these cases via minor modifications. We first give the general strategy in \sect{general} which consists of 3 steps, then discuss the flat spacetime, $d=1$ case, $AdS_3$, general case separately in \sect{examples}.

\subsection{General strategy}\label{sec:general}

\begin{itemize}

\item Step~1:
The states of the form
\be
\Lm{-m_1} \Lm{-m_2} \ldots K_{-n_1} K_{-n_2} \ldots \ket{f},
\label{eq:OCQ_basis1}
\ee
form a complete basis of the Hilbert space ${\cal H}_{\rm OCQ}$, where
\be
K_n \ket{f} = \Lm{n} \ket{f} = 0 \quad (n \geq 1).
\ee
Here, $\ket{f} \in {\cal F}$ is the generalization of DDF states. In the end, DDF states span physical space, so suppose that ${\cal F}$ is known to be unitary. Also, $K_n$ is a generalization of the light-cone operator in Goddard-Thorn's proof. We thus choose $K_n$ which satisfy
\be
[K_m, \Lm{n}]=m K_{m+n}.
\label{eq:light_cone_op}
\ee
However, we do not impose $[K_m, K_n]=0$ unlike Goddard-Thorn's proof; we assume that $K_n$ are nondegenerate. 

\noindent
{\it The outline of the proof:}
The proof begins with mapping the above states to the states of the form 
\be
\LK{-m_1} \LK{-m_2} \ldots K_{-n_1} K_{-n_2} \ldots \ket{f},
\label{eq:OCQ_basis2}
\ee
where $\LK{m} = \Lm{m} - \Lt{m}$. $\Lt{m}$ contains the timelike part and defines a nonunitary $\cX=1$ CFT. $\LK{m}$ defines a Virasoro algebra with $\cK=25$. Then, the proof uses the fact that the Kac determinant for $\cK=25$ is nonvanishing for $\hK>0$. Note that this basis of the Hilbert space \eq{OCQ_basis2} is basically the same as the one in the BRST quantization \eq{basis}.

\item Step~2:
From Step~1, a physical state $\ket{\phi}$ can be written as 
\be
\ket{\phi} = \ket{s} + \ket{k}.
\ee
$\ket{s}$ is a spurious state and $\ket{k} \in {\cal K}$ is a state with no $\Lm{m}$'s. If $\ket{\phi}$ is physical, then $\ket{s}$ and $\ket{k}$ are physical as well. A consequence is that $\ket{s}$ is a null state and $\normb{\ket{\phi}}=\normb{\ket{k}}$.

\noindent
{\it The outline of the proof:}
The proof directly follows from Goddard-Thorn's proof.

\item Step~3:
If $\ket{k}$ is physical, then $\ket{k}=\ket{f}$. Thus,
\be
\normb{\ket{\phi}} = \normb{\ket{k}} = \normb{\ket{f}}.
\ee
So, the space of physical states is unitary if ${\cal F}$ is unitary.

\noindent
{\it The outline of the proof:}
The proof uses the isomorphism of ${\cal K}$ with a $\cX=1$ Verma modules and that the Kac determinant for $\cX=1$ is nonvanishing for $\hX<0$. Since the Kac determinant is nonvanishing, there is no Virasoro primaries in the Verma modules other than $\ket{f}$. 

\end{itemize}

\subsection{Examples}\label{sec:examples}

\begin{itemize}

\item[$\circ$] {\it Flat spacetime or $d \geq 2$ case:}
Originally, Goddard and Thorn \cite{Goddard:1972iy} choose
\be
K_n = k_0 \cdot \alpha_{n},
\ee
where $k_0$ is a specific light-cone vector. (The above proof for Step 1 does not apply to this case, but this step can be shown using the technique of Goddard and Thorn \cite{Goddard:1972iy} or of Thorn \cite{Thorn:1987xy}.) Then, one can immediately conclude that $\normb{\ket{\phi}} = \normb{\ket{f}}$ after Step 2 from $[K_m, K_n]=0$ and $K_n \ket{f} = 0$. However, one needs $d \geq 2$ to define $k_0$.

\item[$\circ$] {\it $d \geq 1$ case:}
As pointed out in Refs.~\cite{Hwang:1991an,Evans:1998qu}, one can alternatively choose
\be
K_n = \alpha^{0}_{n}.
\ee
Then, $K_n$ do not commute, but Step 3 assures $\normb{\ket{\phi}} = \normb{\ket{f}}$. 

\item[$\circ$] {\it $AdS_3$:}
One can choose
\be
K_n = \km{3}{n}.
\ee
For the discrete representation ${\hat{\cal D}}^\pm_j$ and their flowed representations ${\hat{\cal D}}^{\pm,w}_{\tj}$, Step 1 to 3 follow with the bound on the spin $j$ \eq{bound1} since $\hK>0$ only within the bound \cite{Hwang:1991an,Evans:1998qu,Maldacena:2000hw}. On the other hand, for the continuous representations ${\hat{\cal C}}^{\alpha,w}_{\tj}$, this condition on the weight is always valid.
Note that $ J^{3}_{n} \ket{f} = 0$ by construction, so ${\cal F}$ is a subspace of the coset $SL(2,\IR)/U(1)$. The coset has been shown to be unitary within the bound of $j$ in Ref.~\cite{Dixon:1989cg}. 

\item[$\circ$] {\it General cases:}
As one can see, the proof is equally valid as long as one can isolate the timelike part as a $\cX=1$ CFT, and if the basis of the Hilbert space ${\cal H}_{\rm OCQ}$ is written as in Eq.~\eq{OCQ_basis2} or a sum of two Verma modules:
\be
{\cal H}_{\hX,\hK} = {\cal V}(\cX=1, \hX<0) \otimes {\cal V}(\cK=25, \hK>0).
\ee
Note that this is the same as the requirement in the BRST quantization \eq{hilbert}. In general, the operator $K_n$ can be constructed by choosing a free boson basis due to the isomorphism \eq{verma}. 

\end{itemize}


\begin{thebibliography}{999}

\bibitem{Asano:2000fp}
M.~Asano and M.~Natsuume,
``The no-ghost theorem for string theory in curved backgrounds with a flat timelike direction,''
Nucl.\ Phys.\ {\bf B588} (2000) 453
[arXiv:hep-th/0005002].

\bibitem{Asano:2003jn}
M.~Asano and M.~Natsuume,
``The no-ghost theorem in curved backgrounds with a timelike u(1): NSR  string,''
arXiv:hep-th/0303051.

\bibitem{Dixon:1989cg}
L.~J.~Dixon, M.~E.~Peskin and J.~Lykken,
``N=2 superconformal symmetry and SO(2,1) current algebra,''
Nucl.\ Phys.\ B {\bf 325} (1989) 329.

\bibitem{Balog:1988jb}
J.~Balog, L.~O'Raifeartaigh, P.~Forgacs and A.~Wipf,
``Consistency of string propagation on curved space-times: An SU(1,1) based counterexample,''
Nucl.\ Phys.\ B {\bf 325} (1989) 225.

\bibitem{Mohammedi:1989dp}
N.~Mohammedi,
``On the unitarity of string propagation on SU(1,1),''
Int.\ J.\ Mod.\ Phys.\ A {\bf 5} (1990) 3201.

\bibitem{Petropoulos:1989fc}
P.~M.~Petropoulos,
``Comments on SU(1,1) string theory,''
Phys.\ Lett.\ B {\bf 236} (1990) 151.

\bibitem{Bars:1990rb}
I.~Bars and D.~Nemeschansky,
``String propagation in backgrounds with curved space-time,''
Nucl.\ Phys.\ B {\bf 348} (1991) 89.

\bibitem{Hwang:1990aq}
S.~Hwang,
``No ghost theorem for SU(1,1) string theories,''
Nucl.\ Phys.\ B {\bf 354} (1991) 100.

\bibitem{Henningson:1990ua}
M.~Henningson and S.~Hwang,
``The unitarity of SU(1,1) fermionic strings,''
Phys.\ Lett.\ B {\bf 258} (1991) 341.

\bibitem{Hwang:1991an}
S.~Hwang,
``Cosets as gauge slices in SU(1,1) strings,''
Phys.\ Lett.\ B {\bf 276} (1992) 451
[arXiv:hep-th/9110039].

\bibitem{Evans:1998qu}
J.~M.~Evans, M.~R.~Gaberdiel and M.~J.~Perry,
``The no-ghost theorem for AdS(3) and the stringy exclusion principle,''
Nucl.\ Phys.\ B {\bf 535} (1998) 152
[arXiv:hep-th/9806024].

\bibitem{Petropoulos:1999nc}
P.~M.~Petropoulos,
``String theory on AdS(3): Some open questions,''
arXiv:hep-th/9908189.

\bibitem{Maldacena:2000hw}
J.~M.~Maldacena and H.~Ooguri,
``Strings in AdS(3) and SL(2,R) WZW model. I,''
J.\ Math.\ Phys.\  {\bf 42} (2001) 2929
[arXiv:hep-th/0001053].

\bibitem{Pakman:2003cu}
A.~Pakman,
``Unitarity of supersymmetric SL(2,R)/U(1) and no-ghost theorem for  fermionic strings in AdS(3) x N,''
JHEP {\bf 0301} (2003) 077
[arXiv:hep-th/0301110].

\bibitem{Henningson:1991jc}
M.~Henningson, S.~Hwang, P.~Roberts and B.~Sundborg,
``Modular invariance of SU(1,1) strings,''
Phys.\ Lett.\ B {\bf 267} (1991) 350.

\bibitem{Balasubramanian:2002ry}
V.~Balasubramanian, S.~F.~Hassan, E.~Keski-Vakkuri and A.~Naqvi,
``A space-time orbifold: A toy model for a cosmological singularity,''
Phys.\ Rev.\ D {\bf 67} (2003) 026003
[arXiv:hep-th/0202187].

\bibitem{Maldacena:1998uz}
J.~M.~Maldacena, J.~Michelson and A.~Strominger,
``Anti-de Sitter fragmentation,''
JHEP {\bf 9902} (1999) 011
[arXiv:hep-th/9812073].

\bibitem{Seiberg:1999xz}
N.~Seiberg and E.~Witten,
``The D1/D5 system and singular CFT,''
JHEP {\bf 9904} (1999) 017
[arXiv:hep-th/9903224].

\bibitem{Frenkel:1986dg}
I.~B.~Frenkel, H.~Garland and G.~J.~Zuckerman,
``Semiinfinite cohomology and string theory,''
Proc.\ Nat.\ Acad.\ Sci.\  {\bf 83} (1986) 8442.

\bibitem{Goddard:1972iy}
P.~Goddard and C.~B.~Thorn,
``Compatibility of the dual pomeron with unitarity and the absence of ghosts in the dual resonance model,''
Phys.\ Lett.\ B {\bf 40} (1972) 235.

\bibitem{Lian:cy}
B.~H.~Lian and G.~J.~Zuckerman,
``BRST cohomology of the super-Virasoro algebras,''
Commun.\ Math.\ Phys.\  {\bf 125} (1989) 301.

\bibitem{Figueroa-O'Farrill:1989mw}
J.~M.~Figueroa-O'Farrill and T.~Kimura,
``Some results on the BRST cohomology of the NSR string,''
Phys.\ Lett.\ B {\bf 219} (1989) 273.

\bibitem{Figueroa-O'Farrill:1989hu}
J.~M.~Figueroa-O'Farrill and T.~Kimura,
``The BRST cohomology of the NSR string: vanishing and 'no-ghost' theorems,''
Commun.\ Math.\ Phys.\  {\bf 136} (1991) 209.

\bibitem{Griffin:1990fg}
P.~A.~Griffin and O.~F.~Hernandez,
``Feigin-Fuchs derivation of SU(1,1) parafermion characters,''
Nucl.\ Phys.\ B {\bf 356} (1991) 287.

\bibitem{Sfetsos:1991wn}
K.~Sfetsos,
``Degeneracy of string states in 2-D black hole and a new derivation of SU(1,1) parafermion characters,''
Phys.\ Lett.\ B {\bf 271} (1991) 301.

\bibitem{Bakas:1991fs}
I.~Bakas and E.~Kiritsis,
``Beyond the large N limit: Nonlinear W(infinity) as symmetry of the SL(2,R) / U(1) coset model,''
Int.\ J.\ Mod.\ Phys.\ A {\bf 7} (1992) 55
[arXiv:hep-th/9109029].

\bibitem{Kato:2000tb}
A.~Kato and Y.~Satoh,
``Modular invariance of string theory on AdS(3),''
Phys.\ Lett.\ B {\bf 486} (2000) 306
[arXiv:hep-th/0001063].

\bibitem{Maldacena:2000kv}
J.~M.~Maldacena, H.~Ooguri and J.~Son,
``Strings in AdS(3) and the SL(2,R) WZW model. II: Euclidean black hole,''
J.\ Math.\ Phys.\  {\bf 42} (2001) 2961
[arXiv:hep-th/0005183].

\bibitem{Gawedzki:1991yu}
K.~Gawedzki,
``Noncompact WZW conformal field theories,''
arXiv:hep-th/9110076.

\bibitem{Polchinski:rq}
J.~Polchinski, {\it String theory} (Cambridge Univ. Press, Cambridge, 1998).

\bibitem{Thorn:1987xy}
C.~B.~Thorn,
``A detailed study of the physical state conditions in covariantly quantized string theories,''
Nucl.\ Phys.\ B {\bf 286}, 61 (1987).

\bibitem{Pakman:2003kh}
A.~Pakman,
``BRST Quantization of String Theory in AdS(3),''
arXiv:hep-th/0304230.

\bibitem{Hwang:1998tr}
S.~Hwang,
``Unitarity of strings and non-compact Hermitian symmetric spaces,''
Phys.\ Lett.\ B {\bf 435} (1998) 331
[arXiv:hep-th/9806049].


\end{thebibliography}
\end{document}